\documentclass{elsart}

\usepackage{epsfig,bm,epsf,float,amssymb}
\begin{document}

\begin{frontmatter}



\title{Improved Fast Neutron Spectroscopy via Detector Segmentation}
%\title{Demonstration of Enhanced Energy Resolution and Full Energy Deposition Selection for Fast Neutrons Using a Segmented Plastic Scintillator Detector}

\author[sandia]{N.~S.~Bowden\thanksref{thanks1}},
\author[sandia]{P.~Marleau\corauthref{cor1}},
\corauth[cor1]{Corresponding Author. Tel.: +1 925 294 3329.}
\ead{pmarlea@sandia.gov}
\author[sandia]{J.~T.~Steele},
\author[sandia]{S.~Mrowka},
\author[sandia]{G.~Aigeldinger\thanksref{thanks2}},
\author[sandia]{W.~Mengesha\thanksref{thanks3}}

\address[sandia]{Sandia National Laboratories, Livermore, CA~94550, USA}

\thanks[thanks1]{Present Address: Lawrence Livermore National Laboratory, Livermore, CA~94550, USA}
\thanks[thanks2]{Present Address: Touchdown Technologies, Riverside, CA~91706, USA}
\thanks[thanks3]{Present Address: Physical Optics Corporation, Torrance, CA~90501, USA}

\begin{abstract}
Organic scintillators are widely used for fast neutron detection and spectroscopy. Several effects complicate the interpretation of results from detectors based upon these materials. First, fast neutrons will often leave a detector before depositing all of their energy within it. Second, fast neutrons will typically scatter several times within a detector, and there is a non-proportional relationship between the energy of, and the scintillation light produced by, each individual scatter; therefore, there is not a deterministic relationship between the scintillation light observed and the neutron energy deposited. Here we demonstrate a hardware technique for reducing both of these effects. Use of a segmented detector allows for the event-by-event correction of the light yield non-proportionality and for the preferential selection of events with near-complete energy deposition, since these will typically have high segment  multiplicities.

\end{abstract}

\end{frontmatter}

% main text
\section{Introduction}
\label{sec:intro}

Fast neutron detection is a promising technique for Special Nuclear Material (SNM) search and safeguards applications since naturally occurring fast neutrons background rates are relatively low. The ability to distinguish different fast neutrons sources (e.g. fission, ($\alpha$, n) reactions, cosmogenic) via energy spectroscopy would enhance these capabilities.

The function of a spectrometer is to record the energy of an incident particle as accurately as possible.  This first requires that the entire energy of the particle is measured, meaning that the particle must be ``stopped'' within the volume of the detector.  Then, in the case of a scintillation spectrometer, the light yield of the material must be known so that the light pulse amplitude can be converted to the particle energy.  Assuming complete energy deposition and a proportional material response, resolution will be limited only by the statistical variance of the size of the scintillation pulse and its subsequent conversion to an electrical signal. Fast neutron spectrometers based upon organic scintillators often fail in both of these requirements - there is the high likelihood that a fast neutron will leave a detector before it has fully deposited its energy, and there is a non-proportional relationship between deposited energy and the observed scintillation light.

Let us consider the second point more fully.  Both gamma rays and neutrons typically undergo several interactions within a detection material before losing all of their energy.  In the case of gamma rays, energy is transferred to electrons; in the case of neutrons in organic scintillators, most of the energy is transferred to protons.  These secondary charged particles lose energy via ionization in the detector material, which ultimately results in pulses of visible light.

The conversion of this dissipated energy to a light pulse is described by a function $L(E_i)$, where $E_i$ is the energy lost by the incident particle during the $i$th interaction in the detection medium. It has been known for decades that both inorganic~\cite{Engelkmeir} and organic~\cite{Birks} materials have a non-proportional energy response. Therefore, if only the total emitted light is recorded, an incorrect particle energy will be inferred.

To see this consider a situation where the light yield function $L$ is linear in energy. The total light observed due to the incident particle will be
\begin{equation}
\sum_{i} L(E_i) \propto  \sum_{i} E_i ,
\label{eq:prop_yield}
\end{equation}
i.e. the observed scintillation light pulses will be proportional to the energy lost by the incident particle in the detection medium. However, in the case of a nonproportional light yield
\begin{equation}
\sum_{i} L(E_i) \neq  L\left(\sum_{i} E_i\right) ,
\label{eq:nprop_yield}
\end{equation}
and the magnitude of the observed light pulse will depend on the microscopic detail of the way in which the incident energy is divided between the $i$ interactions.

In many circumstances this non-linearity dominates statistical contributions to the energy resolution. For example, the energy resolution of NaI(Tl) at $667$~keV would be expected to be about 3\% if photo-electron statistics dominated - the observed value of close to 7\% is caused by the non-linearity of $L(E)$. One approach to overcoming this limitation is to produce materials with linear $L(E)$ - this is a path that has recently yielded great success for gamma-ray scintillators,  with the production of Ce doped scintillators like LaBr(Ce) which have very flat light yield curves~(see, e.g., \cite{Dorenbos}).

In the case of organic scintillator neutron spectrometers, it has long been recognized that the non-linear sum of the scintillation light from successive recoil protons is an important energy resolution limiting effect. The resulting smeared energy response has been measured for monoenergetic input neutron beams (see for example, \cite{Aoyama}) and/or estimated via Monte-Carlo codes, and then corrected using statistical unfolding.

\begin{figure}[tb]
\centering
\includegraphics*[width=3in]{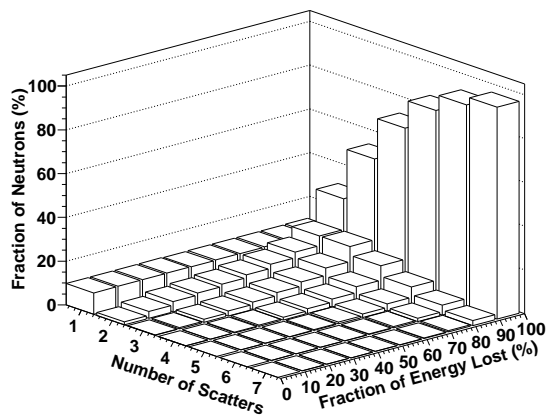}
\caption{The fraction of energy lost by $5$~MeV neutrons as a function of the number of proton scatters undergone in a $25.4$~cm x $10.16~$cm x $10.16~$cm plastic scintillator bar. After 3 scatters 71\% of neutrons have lost more than 80\% of their energy.} \label{fig:Eperscat}
\end{figure}

Another approach, one that relies on neither materials development nor statistical deconvolution, is to attempt to resolve each individual energy deposition via detector segmentation~\cite{Abdurashitov}. One would seek to observe only one interaction per detector segment so that the nonlinear light yield function can be correctly inverted for each recoil proton. This approach does not work for inorganic scintillator crystals used for gamma rays, as much of the energy loss occurs via numerous delta rays that would be practically impossible to resolve~\cite{Mengesha}.

Detector segmentation may have another benefit. Due to the relatively small (n,p) elastic scattering cross section, there is a considerable likelihood that a neutron will leave a detector before all of its energy is deposited. However, in a segmented device, neutrons that are observed to undergo many scatters have a high likelihood of having deposited most of their energy (Fig.~\ref{fig:Eperscat}). Therefore a cut that chooses events that interact in many segments will preferentially select near-full energy depositions. For example, an event sample produced by a segment multiplicity cut requiring more than three segments to be hit would contain 71\% of events with energy deposition greater than 80\%.

There are other techniques used to account for incomplete energy depositions in organic scintillator fast neutron spectrometers. This effect is included in statistical unfolding codes, or can be addressed on an event-by-event basis via the capture-gating technique (see, e.g. \cite{Czirr,Feldman,Aoyama}). In this last case the inclusion of a capture agent or separate capture detector adds to the complexity of the design, readout, and data analysis of the system. For example, when using Li or B capture agents in organic scintillator, the resulting capture signal is heavily quenched and thus difficult to observe in high background environments. The combination of improved energy resolution and full-energy deposition selection may therefore make detector segmentation an attractive option. To demonstrate both of these concepts we have designed, constructed and tested a segmented plastic scintillator detector.

\section{Design of the segmented plastic scintillator detector}
\label{sec:design}

BC-408 organic plastic scintillator was selected for this initial demonstration for ease of handling and construction. The total plastic scintillator volume utilized is $\approx 2500$~cm$^3$ ($\approx 25$~cm x $10~$cm x $10~$cm).

\begin{figure}[tb]
\centering
\includegraphics*[width=3in]{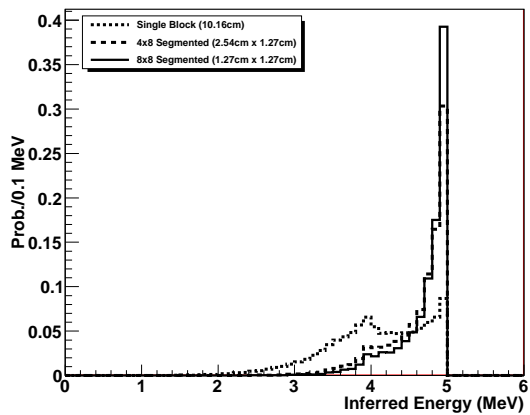}
\caption{The energy inferred from a plastic scintillator detector for $5$~MeV neutrons that lose all of their energy in the detector. As the number of segments is increased and their size decreased, more individual proton recoils are resolved, allowing a more accurate energy inference to be made.} \label{fig:E_enhance}
\end{figure}

The size of the segments was chosen based upon modeling using GEANT4~\cite{GEANT}. In practice, a compromise must be struck between performance on the one hand and cost and complexity on the other. Our selection of a segment cross section of $1.27$~cm x $2.54$~cm is a reasonable compromise, in that it should result in a considerable resolution improvement (Fig.~\ref{fig:E_enhance}), it is compatible with available Photomultiplier Tubes (PMTs), and results in a reasonable number of channels to read out.

\begin{figure}[tb]
\centering
\includegraphics*[width=3in]{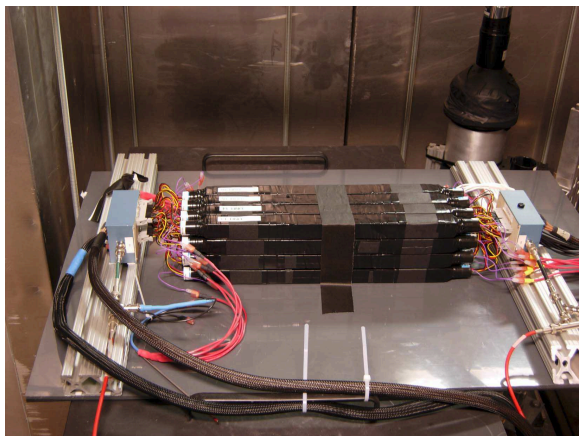}
\caption{A photograph of the array of plastic scintillator segments.} \label{fig:detector}
\end{figure}

Hamamatsu R1548 double anode PMTs were used to detect scintillation light produced in the segments - the photo-cathode of each channel matches the segment cross section selected. Our design therefore consisted of 32 plastic scintillators bars, 1.27cm x 2.54cm x 25.4 cm, each wrapped in white Teflon tape to maximize light collection, assembled into a $4~$x$~8$ array (Fig.~\ref{fig:detector}). PMTs were coupled to both ends of the scintillator segments with optical grease, for a total of $64$~readout channels.

Using a PMT on each end of a bar improves light collection uniformity. Only those energy depositions for which the signal amplitudes from each end of one bar were approximately equal were used in the analysis. This necessarily constrained the interaction region to near the detector array center. This was done to ensure the cleanest possible demonstration of correction for non-linearity from positional effects. In the future, a more detailed positional calibration to determine interaction position along the bars could be used to increase the active volume of the device.

\section{Signal Processing and Data Collection:}
\label{sec:DAQ}

\begin{figure}[tb]
\centering
\includegraphics*[width=3in]{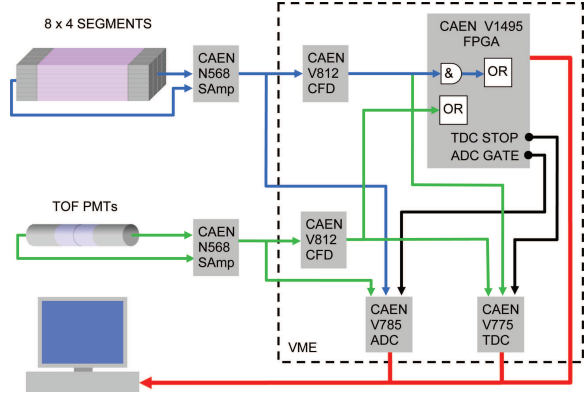}
\caption{Segmented neutron detector electronics.  The time of flight (TOF) capability is included. From the photomultiplier (PMT) pulses, constant fraction discriminators (CFD) generated timing/logic pulses, which were processed by the FPGA module. The Spectroscopy Amps and capture CFD were implemented in NIM. The analog to digital converter (ADC), time to digital converter (TDC), other CFDs and the FPGA were implemented in VME. Thicker lines denote multiple signals. (ORTC=ORTEC).} \label{fig:daq}
\end{figure}

When a neutron interacts in the segmented detector the majority of its energy is lost in the first few scatters, which occur within a few nanoseconds.  Since this time is very short compared to the response time of the Gaussian shaping/peak sensing pulse analysis used, readout was greatly simplified by reading the response of all segments simultaneously.  The segments containing the interactions of interest were later identified offline during data analysis.  A schematic of the segmented detector instrumentation is shown in Fig.~\ref{fig:daq}.  The spectroscopy amplifiers were implemented in NIM and all other modules in VME.  The VME and NIM systems were controlled and data were collected by a LabVIEW program running on a Windows computer.

The 64~PMTs contained within the 32~Hamamatsu R1548 units were all biased at $1850$~V using a single Stanford PS350 high voltage supply.  Pulses from the PMTs were amplified using CAEN~N568LC 16~Channel programmable spectroscopy amplifiers with a shaping time of $1~\mu$s.  The amplitudes of the shaped PMT pulses were digitized using~CAEN V785 peak sensing ADCs.  The spectroscopy amplifiers also provided fast, unshaped, fixed gain timing outputs used to drive CAEN~V812 constant fraction discriminators (CFD) whose outputs were subsequently used to generate timing/logic signals.  Because of the common PMT bias voltage, the spectroscopy amplifier gains required adjustment to compensate for the inevitable gain differences between PMTs.  Because of the large number of channels, gain calibration using gamma sources was automated by employing a rudimentary Compton edge finder algorithm in the control software with only enough resolution as required by the knowledge that more accurate offline calibrations were to be performed as described in Sec.~\ref{sec:calibration}.

When operating in normal mode, the custom logic implemented in a CAEN~V1495 FPGA module generated a $1.5~\mu$s wide ADC gate whenever there were coincident outputs from the two PMTs in any segment above CFD threshold.  The FPGA module was also responsible for enforcing module conversion dead-times, synchronizing module data readout and providing scalars and timers for use in rate measurements.

In order to better characterize the detector, a time-of-flight (TOF) scatter system was incorporated to enable determination of the incident neutron energy (Sec.~\ref{sec:TOF}). A CAEN~V775 time to digital converter (TDC) was added to provide the time measurements.  In TOF mode, the FPGA was programmed to wait for a coincidence between a deposition above threshold in at least one TOF PMT and both PMTs of at least one segment detector within a coincidence window of $175$~ns.  When this coincidence occurred, the FPGA would generate both a TDC stop pulse and a $1.5~\mu$s wide TOF/segment ADC gate.  The TDC start pulses were supplied earlier and directly by the CFDs for the TOF scatter and segment detectors.  The segment detector start pulses were derived from PMTs all located on the same end of the detector block.  The TDC was setup to give $75$~ps per bit resolution, giving a full scale range of $288$~ns.  This provided more than adequate resolution with respect to path length uncertainties for neutron energy computations and enough range to work with the latency of the trigger logic.  Since only $64$~channels of electronics were available, the TOF system was added by removing energy readout from both ends of one scintillator bar and timing readout from this and another bar and replacing them with readouts of two plastic scintillator TOF detectors.  The segment detector bars chosen for removal were from the side opposite the source, since such segments were the least likely to be involved in a significant way in the multiple recoil interactions of greatest interest.

\section{Data Analysis}
\label{sec:analysis}

The data for one event consisted of $64$~ADC values, followed by $32$~TDC values, followed by FPGA codes pertaining to the conditions of the trigger and the state of the data acquisition.  Each detector bar was assigned two ADC channels (one for each PMT on either end of the bar) and a single TDC channel with input from the discriminated output of one PMT (same end for all bars).

In order to determine which detector segments were involved in a single triggered event, a software ADC threshold was applied to each ADC channel.  The thresholds were determined independently for each ADC channel by requiring that the ADC value be at least 50~units above the pedestal value.  Pedestal values were determined by histogramming the entire data set for each channel and searching for the lowest value peak in the distribution.  This procedure resulted in each bar detector having a different lower energy threshold, but allowed for the lowest possible energy threshold for the segmented detector as a whole.  The thresholds in electron equivalent energy ranged from $20$~keV to $70$~keV with an average of ~$35$~keV.

Data was further reduced by requiring that both ends of each bar detector have ADC values above the software threshold and that the ratio of the difference to the sum of the pedestal subtracted ADC values be less than~$0.33$ (i.e. neither end was allowed to have less than 50\% the value of the other).  These requirements were implemented to accept only those events in which all scatters in the block detector deposited energy primarily in the central regions of each bar.

Lastly, any event that included any number of ADC channels at, or near, saturation (greater than an ADC value of~$3800$) was rejected.  This final condition removed events with a muonic component.

\subsection{Calibration}
\label{sec:calibration}
Once it was determined that a detector segment contained a scatter associated with the event, the energy deposited was determined by converting the average of the ADC values of the two ends of the segment into an electron equivalent energy.

The parameters of this linear conversion were determined independently for each detector segment from calibrations using gamma sources.  To achieve this, the average ADC values of the two ends filled a separate 1 dimensional histogram for each segment. The ADC value of the Compton edge for several mono-energetic gamma ray energies was found using a gaussian fit to pulse height spectrum near the Compton edge and solving for the edge location using the procedure suggested in \cite{Dietze}.

%The Compton edge for each mono-energetic gamma was found by fitting the resulting spectrum, finding the peak near the Compton edge and solving for the ADC value at 70\% of the peak height.  This location has been identified as the Compton edge energy in detectors of similar geometry \ref{Dietze}.  This was done for three gamma energies:  The $^{22}$Na $0.511$~MeV and $1.275$~MeV lines and $^{137}$Cs $0.662$~MeV line.

%The ADC value of the 70\% peak versus the energy of the Compton edge were then fit to a straight line. The slope and intercept of the fit provided the conversion from ADC value to electron equivalent energy.

Assuming that the target particle in each scatter was a proton, the electron equivalent energy could then be corrected for the effects of quenching by applying the following non-linear function \cite{Aoyama,Knoll};
\begin{equation}
E_p = F(E_{ee}) =  \left(\frac{E_{ee}}{3}\right)^{2/3} ,
\label{eq:ly}
\end{equation}
where $E_{ee}$ is the electron equivalent energy deposited and $E_p$ is the actual energy imparted to the proton in the scatter. This generic representation of the response of organic scintillator to protons will suffice for this demonstration. However, for optimal performance, this response should be carefully calibrated for the particular material to be used.

\subsection{Proton Quenching Corrections}
\label{sec:quenching}

Once the electron equivalent energy was determined for each scatter within an event, the total energy deposited in the segmented detector was determined in two ways.  First, a baseline, non-segmented block detector was approximated by summing the electron equivalent energies for all scatters and then applying the quenching correction;
\begin{equation}
E_{block} =  F\left(\sum_i E_{ee_i}\right).
\label{eq:block}
\end{equation}
This essentially erased the advantages of segmentation by effectively integrating the light in all scatters in a manner analogous to the way a single large volume detector would sum the scintillation light from all scatters in a single event.

This was then compared to the calculation of the total energy deposited utilizing segmentation.  Here, each scatter was corrected for the effects of quenching before the summation:
\begin{equation}
E_{seg} =  \sum_i F(E_{ee_i}).
\label{eq:seg}
\end{equation}
Because the correction function $F(E_{ee_i})$ was non-linear, these two methods of determining the total energy, $E_{block}$ and $E_{seg}$ would not be identical unless the event contained only a single scatter.

\section{Neutron Initial Energy Determination via Time of Flight}
\label{sec:TOF}

In order to compare the accuracy of both methods of determining the total energy of a neutron interaction, a source of neutrons of known energy was needed.  To achieve this, we implemented a TOF detector consisting of two $2.54$~cm thick by $2.54$~cm diameter plastic scintillators coupled to PMTs.  A 32~mCi AmBe neutron source was placed $\approx 15$~cm from the center of the TOF detector, which in turn was placed $90$~cm from the segmented plastic scinitllator detector.  A trigger was implemented by requiring that both the segmented block detector and the TOF detector have energy deposited in them above threshold within a coincidence window of $175$~ns duration.

\begin{figure}[tb]
\centering
\includegraphics*[width=3in]{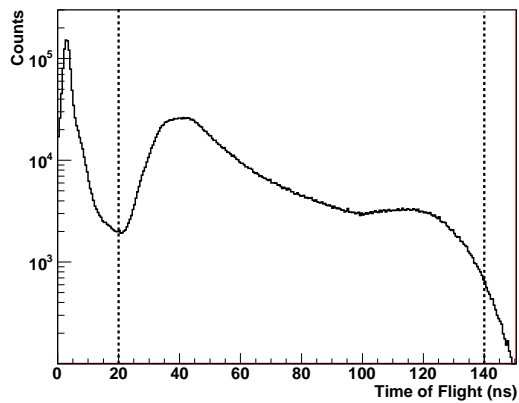}
\caption{TOF distribution for coincident events.  Neutrons and gamma rays are separated by the cut shown at $20$~ns. } \label{fig:TOF_result}
\end{figure}

\begin{figure*}[tb]
\centering
\includegraphics*[width=6in]{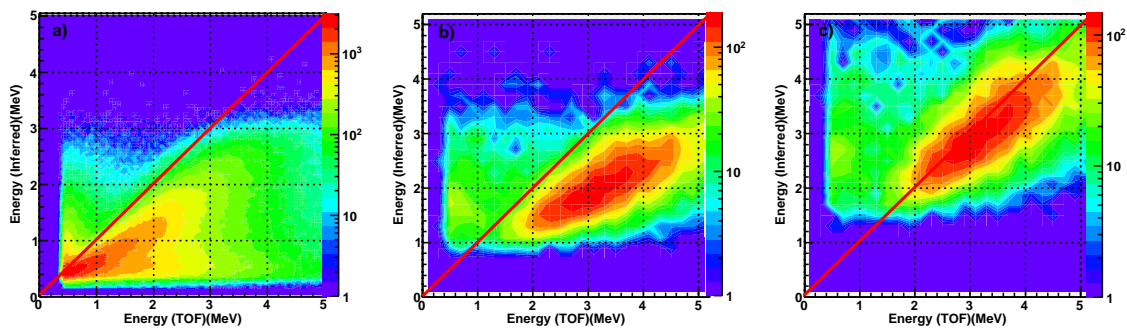}
\caption{Comparisons of the incoming neutron energy ($E_{TOF})$ and that inferred ($E_{Inferred}$) by: (a) treating the detector as a single block; (b) requiring 3 or more resolved scatters in the detector and therefore preferentially selecting near full energy depositions; and (c) requiring 3 or more resolved scatters and applying the non-proportionality correction to each resolved scatter.} \label{fig:All_result}
\end{figure*}

The TOF of a scattered neutron was then determined by the difference between the scattering times in the scatter detector and the first scatter in the block detector using TDC data.  Any TOF less than $20$~ns was rejected in order to eliminate gamma rays (background or from the AmBe source) that may have Compton scattered in both detectors.  Any TOF greater than $140$~ns was also rejected both because neutrons slower than this were most likely to be below the energy threshold of the block detector and because this shortened the coincidence window to reduce accidental "false" coincidences  (Fig.~\ref{fig:TOF_result}).  The energy ($E_{n_i}$) of each neutron incident on the segmented block was then determined by
\begin{equation}
E_{n_i} =  \frac{m}{2} \left(\frac{d}{t_{TOF}}\right)^2,
\label{eq:tof}
\end{equation}
where $m$ is the neutron mass, $d$ is the distance between scatters and $t_{TOF}$ is the measured time of flight.  The TOF cuts discussed above limited the energy range of incident neutrons to between $\approx220$~keV and $\approx10$~MeV.

\section{Results}
\label{sec:results}

To assess the performance of segmented detector for both full energy deposition selection and fast neutron spectroscopy, we compare the energy inferred by this device, $E_{Inferred}$ with that determined by the TOF method, $E_{TOF}$. We begin by comparing the energy inferred without using the detector segmentation (Fig.~\ref{fig:All_result}a). On average, the energy inferred from the detector is considerably lower than that of the incoming neutron. This difference is primarily caused by incomplete energy depositions, as well as by the non-proportional light yield of the material.

Next we utilize the detector segmentation to preferentially select neutrons that have deposited a large fraction of their energy in the detector (Fig.~\ref{fig:All_result}b). The proportionality between the incoming neutron energy and that inferred is considerably improved, although the inference is still systematically low.

Finally, we utilize the detector segmentation to properly account for the scintillator light yield non-proportionality as described in Sec.~\ref{sec:quenching} (Fig.~\ref{fig:All_result}c). The mean inferred energy approaches that of the incident neutron, as determined by the TOF method. The segmentation of the detector has allowed us, on an event-by-event basis, to select events in which a neutron has deposited most of its energy, and to more accurately infer that energy of that neutron.

\begin{figure}[tb]
\centering
\includegraphics*[width=3in]{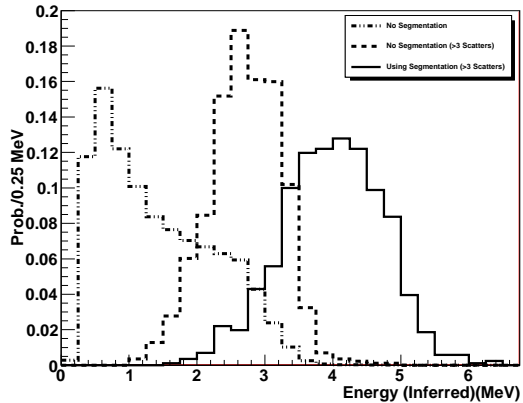}
\caption{Inferred energies for neutrons with $E_{TOF}$ between $4.75$ and $5.25$~MeV. All histograms are scaled to have the same area. Using segmentation to select near complete energy depositions (dashed line) and to fully correct for the nonproportional light yield from each proton recoil (solid line) yields a much improved energy inference on an event by event basis than treating the detector as a single solid block (dot-dashed line).} \label{fig:5MeV_result}
\end{figure}

This can also be seen in Fig.~\ref{fig:5MeV_result}, where we examine only events with incident neutron energies within $0.25$~MeV of $5$~MeV. The average inferred energy approaches $5$~MeV as first full energy deposition selection is applied, and then as the scintillator non-proportionality is corrected for. These last two curves can be compared to the prediction of the inferred energy spectrum based upon a neutron transport simulation in Fig.~\ref{fig:E_enhance}.

It can be seen that our current implementation of full energy deposition selection preferentially selects events in which the incoming energy is fairly evenly shared amongst a few scatters - there are few events near the incoming energy in the dashed curve in Fig.~\ref{fig:5MeV_result}. This can be explained by our fairly high segment thresholds, which result in an event with a single large energy deposition not meeting the multiplicity criteria. Lower thresholds, and an analysis threshold applied to the detector as a whole, rather than segment by segment, could improve this situation.

Also, the inferred neutron energy is still, on average, lower than the incoming neutron energy. This is due to a number of reasons. First, selecting three or more scatters does not guarantee full energy deposition (Fig.~\ref{fig:Eperscat}). Requiring greater multiplicity would bring the average inferred energy closer to incident energy, at the cost of lower efficiency. Second, if the energy deposited in a segment does not result in enough light to exceed our analysis threshold for that segment, that energy is not included in our inference. This will always reduce the average inference and can be improved upon via lower thresholds or a detector wide thresholding scheme as described in the previous paragraph. Finally, one can never resolve all individual scatters, meaning the the non-proportionality correction is not perfect. On average, this results is a slightly lower inferred energy, as can be seen in Fig.~\ref{fig:E_enhance}.

\section{Conclusion}
\label{sec:conclusion}

The addition of segmentation to an organic scintillator detector has been shown to be advantageous for fast neutron spectroscopy.  By resolving multiple proton recoils within a detector, we are able select events for which there is high likelihood of full energy deposition having occurred and to correct for the non-proportional light yield of the detector material.

The results presented here could be improved upon in several respects. Lower segment thresholds and/or a thresholding scheme based upon the total light observed within the device would yield greater efficiency. Use of an organic liquid scintillator would allow for gamma-ray rejection via Pulse Shape Discrimination, which would enhance the applicability of this technique in high background environments.

The selection of full energy deposition by this means is in some respects simpler to implement than neutron capture gating, and it has the added benefit of allowing for the correction of the inherent non-linearity of organic scintillators. We expect this technique to have similar efficiency to capture gating, since this later method is sensitive to the same subset of incident neutrons - those that happen to lose all of their energy in the detector.

\section*{Acknowledgements}
%\begin{ack}
%\section{Acknowledgements}
%\label{sec:acknowledgements}

This work was supported by Laboratory Directed Research and Development (LDRD) at Sandia National Laboratories.
Sandia is a multiprogram laboratory operated by Sandia Corporation, a Lockheed Martin Company, for the United States Department of Energy¢s National Nuclear Security Administration under Contract DE-AC04-94AL85000.
%\end{ack}
%\end{Acknowledgements}

\end{document}